\documentclass[aps,prl,twocolumn,showpacs,groupedaddress]{revtex4}
\usepackage{amsbsy,amssymb,amsmath,bm} \usepackage{graphicx}

\input{epsf}
\begin{document}

\title{Collective Cooper-pair transport in the insulating state of Josephson junction arrays}
\author{M. V. Fistul$^1$, V. M. Vinokur$^2$, and T. I.  Baturina$^{3,2}$}
 \affiliation{$^1$Theoretische Physik III,
Ruhr-Universit\"at Bochum, D-44801 Bochum, Germany\\
$^2$Material Science Division, Argonne National
Laboratory, Argonne, Ill. 60439 USA\\
$^3$Institute of Semiconductor Physics, 630090, Novosibirsk,
Russia
}
\date{\today}

\begin{abstract}

We investigate collective Cooper-pair transport of one- and
two-dimensional Josephson junction arrays in the insulating state.
We derive an analytical expression for the current-voltage
characteristic revealing thermally activated conductivity at small
voltages and threshold voltage depinning.  The activation energy and
the related depinning voltage represent a dynamic Coulomb barrier
for collective charge transfer over the whole system and scale with
the system size. We show that both quantities are non-monotonic
functions of magnetic field. We propose that formation of the
dynamic Coulomb barrier as well as the size scaling of the
activation energy and the depinning threshold voltage, are
consequences of the mutual phase synchronization. We apply the
results for interpretation of experimental data in disordered films
near the superconductor-insulator transition.

\end{abstract}

 \maketitle

Recent experimental studies of the supercon\-ductor-insulator
transition (SIT) (see~\cite{Girvin-rev} for a review) in thin
disordered superconducting films proved formation of a collective
insulating state exhibiting thermally activated Arrhenius-like
conductivity at low biases~\cite{Shahar-act,Baturina,Ovadyahu} and
the threshold voltage depinning~\cite{Shahar-Coll,Baturina}
behavior. Below the depinning voltage, $V_T$, a film falls into  a
zero-conductivity phase and abruptly switches to a finite
conductance regime when the bias achieves $V_T$. The discovery of
the novel phase existing in a narrow window of disorder strength
near SIT poses a quest for a comprehensive theory.

Josephson junction arrays serve as a perfect testing ground for SIT
studies (see, e.g.,~\cite{Efetov,JJA,Bradley,Mooij,Haviland}). A
salient similarity of the voltage threshold behavior in SC
films~\cite{Shahar-Coll,Baturina} and the voltage depinning in
one-dimensional Josephson arrays~\cite{Haviland} suggests an
intimate relation between these systems.  Further parallel appears
from the striking observations of voltage threshold $V_T$ dependence
on the array length in~\cite{Haviland}, the sample size dependent
activation energy, $k_BT_0$, observed in~\cite{Ovadyahu}, and the
connection between $V_T$ and $T_0$ revealed in~\cite{Baturina}. An
advantage of a Josephson array as a model system is that it offers a
straightforward theoretical description of the current-voltage
characteristics, which is what measured in all the major
experimental studies of SIT.  In this Letter we develop a theory of
the collective transport of large Josephson junction arrays in the
insulating state and apply our results for interpretation of
experimental data on SIT.

\begin{figure}[tbp]
\includegraphics[width=2.2in]{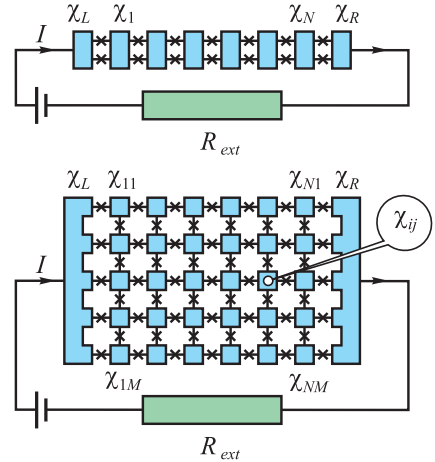}
\caption{(Color online) Sketch of the considered array geometries.
An external current $I$ is injected from the left through the
electrode having the superconducting phase $\chi_{\scriptscriptstyle
L}$ and extracted through the right electrode with the phase
$\chi_{\scriptscriptstyle R}$.  Upper panel: One-dimensional array
of $N$ superconducting islands (squares) connected by two Josephson
junctions (crosses) to neighbors corresponding to experimental
system of~\cite{Haviland}. Lower panel: Two-dimensional $M\times N$
Josephson junction array.  } \label{circuit}
\end{figure}

The current-voltage characteristics of Josephson systems in an
insulating state were discussed in a single
junction~\cite{Ingold,Tinkham,LikhAverin} and two
junction~\cite{Matveev,Zorin} systems.  Each junction is
characterized by the Josephson coupling energy, $E_{J0}=\hbar
I_c/2e$, where $I_c$ is the Josephson critical current, and by
charging energies $E_c$ related to inter-island capacitance and
$E_{c0}$ associated with capacitance to ground, $C_0$~\cite{note1}.
We consider an insulating state, where charging energies
$E_c,E_{c0}\gg E_J$.  We assume further the superconducting gap
$\Delta>E_c$. This implies that the transport is mediated by the
thermally activated motion of the Cooper pairs~\cite{LopVin}.  We
show that in the regular and disordered arrays a collective current
state develops. This state is characterized by the energy gap,
$\Delta_c$, stemming from the coherent Coulomb blockade effect, i.e.
the Coulomb blockade involving all junctions and extending over the
whole system. We derive a low bias $I$-$V$ dependence:
\begin{equation}
I\propto\exp\bigg[-\frac{(\Delta_c-eV)^2}{2\Delta_ck_BT}\bigg].
\label{IV}
\end{equation}
Eq.(\ref{IV}) reveals that there are two dynamic regimes: first,
thermally activated charge transfer with the resistance
\begin{equation}
R\propto\exp[\Delta_c/(2k_BT)]\label{activation}
\end{equation}
at $eV\ll\Delta_c$ and, second, threshold behavior at $V\approx
V_T\simeq\Delta_c$, where the activated conductivity turns to a
finite non-activated transport. We find that in a regular 1D array
$\Delta_c\simeq E_cL/d$, while in a 2D array $\Delta_c\simeq
E_c\ln(L/d)$, where $L$ is the array length and $d$ is the size of
the elemental cell of the array. We show that at the depinning
voltage threshold in disordered 2D systems the current breaks
through the system along the first percolative path connecting the
leads. As a result, the depinning transition acquires an
one-dimensional character, and the depinning voltage $V_T$, scaling
linearly with the sample length, much exceeds the corresponding low
bias activation energy. We demonstrate, finally, that the magnetic
field perpendicular to the film gives rise to a non-monotonic
$\Delta_c(B)$ dependence in an excellent agreement with the
experimental data on both 1D artificial Josephson
arrays~\cite{Haviland} and superconducting films near
SIT~\cite{Shahar-Coll,Baturina}.

Let us consider $N\times M$ superconducting islands comprising a
two-dimensional array closed by a small (as compared to the quantum
resistance for Cooper pairs $R_{CP}=h/4e^2\simeq 6.45$\,k$\Omega$)
external resistance, $R_{ext}$, see Fig. 1. We assign the
fluctuating order parameter phase $\chi_{ij}(t)$ to the $\{i,j\}$-th
superconducting island (see Fig. 1). The phases of the left- and
right leads, $\chi_{L}(t)$ and $\chi_{R}(t)$, respectively, are
fixed by the dc voltage $V$ across the array:
\begin{equation}
\chi_{R}-\chi_{L}=2eVt/\hbar+\psi(t)~, \label{classicalphases}
\end{equation}
where $\psi(t)$ describes fluctuations in the leads.  We single out
the leftmost, $i=1$, and rightmost, $i=N$, columns of islands
directly coupled to leads and represent the array Hamiltonian in a
form:
\begin{eqnarray}\nonumber
H=H_0+H_{int}+
\frac{\hbar^2}{8E_c}\sum_{j=1}^M(\dot{\chi}_{1j}(t)+\dot{\chi}_{Nj}(t))^2-\\
-2E_J\sum_{j=1}^M\cos\left[ \frac{\chi_{1j}(t)+\chi_{Nj}(t)}{2}\right]\times\\
\nonumber \times \cos\left[
\frac{2eVt/\hbar+\psi(t)+\chi_{1j}(t)-\chi_{Nj}(t)}{2}\right].
\label{Hamiltonian}
\end{eqnarray}
Here
\begin{multline}
H_0=\sum_{\langle ij,kl\rangle}\Bigl[\frac{\hbar^2}{4E_c}(\dot{\chi_{ij}}-\dot{\chi_{kl}})^2-E_J\cos(\chi_{ij}-\chi_{kl})\Bigr]\\
+\sum_{ij}\frac{\hbar^2}{4E_{c0}}\dot{\chi}_{ij}^2~, \label{ham}
\end{multline}
the brackets $\langle ij,kl\rangle$ denote summation over the pairs
of adjacent junctions, and the last term in (\ref{ham}) represents
the self-charge energies of superconducting islands. The $H_{int}$
term in (4) describes coupling of phases on the leads  to the
thermal heat bath~\cite{Ingold}.

The dc Josephson current through the array is
\begin{eqnarray}\nonumber
I_s(V)&=&I_c\lim_{\tau \rightarrow \infty} \frac{1}{\tau}\int_0^\tau
dt \sum_{j=1}^M\bigg\langle
\cos\left[ \frac{\chi_{1j}(t)+\chi_{Nj}(t)}{2}\right]\times\\
&\times&\sin\left[
\frac{2eVt/\hbar+\psi(t)+\chi_{1j}(t)-\chi_{Nj}(t)}{2}
\right]\bigg\rangle, \label{current}
\end{eqnarray}
where the brackets $\langle ...\rangle$ stand for an averaging over
thermal fluctuations in the leads and quantum mechanical averaging
over phases of internal junctions ($\chi_{ij}(t)$) and over the
variable $\phi_j=(\chi_{1j}+\chi_{Nj})/2$. We construct the
time-dependent perturbation theory with respect to small parameter
$E_J/E_c$, similarly to the case of the Cooper pair two-junctions
transistor~\cite{Matveev,Zorin}, omitting the last term in
(\ref{ham}) since in most experiments $C\gg C_0$, and thus $E_c\ll
E_{c0}$.  In the first order one finds:
\begin{equation}
\big\langle\cos \phi_j\big\rangle =\frac{E_J}{2E_c}\cos\left[
\frac{2eVt/\hbar+\psi(t)+\chi_{1}(t)-\chi_{N}(t)}{2}\right].
\end{equation}
In the second order, using the approach developed in \cite{Ingold},
one arrives at:
\begin{equation}
I_s(V)=MI_c\frac{E_J}{\hbar}\bigg(\frac{E_J}{2E_c}\bigg)^2 \Im m
\int_0^{\infty}dt e^{-t\delta/\hbar}K(t)e^{i\frac{2eVt}{\hbar}},
\label{current2}
\end{equation}
where $\delta=4e^2R_{ext}k_BT$ reflects the Gaussian character of
the current noise in the leads due to thermal fluctuations
\cite{Ingold,KovFistUst}.  The correlation function  of internal
phases is defined as
\begin{equation}
K(t)=\langle \exp
(i[\chi_{1j}(t)-\chi_{1j}(0)-\chi_{Nj}(t)+\chi_{Nj}(0)
])\rangle_{H_0}. \label{corrfunct}
\end{equation}
In the two-junction system (single Cooper pair transistor),
$\chi_{1j}\equiv\chi_{Nj}$, $K(t)\equiv 1$, and we recover the
results of~\cite{Zorin}. In the zero-approximation one neglects the
Josephson coupling inside the array, and $K(t)$ can be found in a
closed form as an analytical continuation of $K(\tau)$, where $\tau$
is the imaginary time:
\begin{multline}
K(\tau)=\int D[\chi_{ij}]\exp(
i[\chi_{1j}(\tau)-\chi_{1j}(0)-\chi_{Nj}(\tau)+\chi_{Nj}(0)])\\
\exp\bigg({-\frac{\hbar}{4}\int_0^{\frac{\hbar}{k_BT}}d\tilde\tau
\bigg[ \sum_{\langle
ij,kl\rangle}\frac{[\dot{\chi}_{ij}(\tilde\tau)-\dot{\chi}_{kl}(\tilde\tau)]^2}{E_c}
-\sum_{ij}\frac{[\dot{\chi}_{ij}(\tilde\tau)
]^2}{E_{c0}}\bigg]}\bigg)\,. \label{korrfunctCalc}
\end{multline}
Expanding phases $\chi_{ij}(\tilde\tau)$  over the Matsubara
frequencies $\omega_m=2\pi k_B Tm/\hbar$ as $
\chi_{ij}(\tilde\tau)=\sum\exp(i\omega_n
\tilde\tau)\chi_{ij}(\omega_m)\label{expansionMatsfr}, $ and going
over to charge representation, $\chi_{ij}(\omega_m)=n_{ij}(2E_c k_B
T/(\hbar\omega_m)^2)[\exp(-i\omega_m \tau)-1]$, with $n_{ij}$ being
the number of Cooper pairs localized on the $\{ij\}$-th island, one
eventually obtains the correlation function $K(t)$:
\begin{equation}
K(t)= \exp{\big(-2\Delta_ck_B
Tt^2/{\hbar}^2-2i\Delta_ct/\hbar\big)}\label{korrfunctlargeT}~,
\end{equation}
where $\Delta_c$ is the barrier for the {\it Cooper pair
propagation} through the whole system defined by the relation
\begin{multline}
\exp\bigg(-\frac{\Delta_c}{k_BT}\bigg)= \int D[n_{ij}]\exp
\frac{E_c}{k_BT}\bigg[i(n_{1j}-n_{Nj})-\\
- \sum_{\langle ij,kl\rangle}\frac{1}{2}(n_{ij}-n_{kl})^2
-\sum_{ij}\frac{E_c n^2_{ij} }{2E_{c0}}\bigg] \label{Coulombgap}.
\end{multline}

Importantly, when deriving
Eqs.\,(\ref{korrfunctCalc}-\ref{Coulombgap}), the non-zero windings
numbers $W_{ij}=[\chi_{ij}(\hbar/k_BT)-\chi_{ij}(0)]/(2\pi)$ were
neglected.  This holds only at not very low temperatures,
$T>E_c/(\pi k_B)$. At $T=E_c/(\pi k_B)$ a
Berezinskii-Kosterlitz-Thouless-like transition into a
super-insulating state with the resistivity
$R\propto\exp\{(\Delta_c/E_c)\exp[E_c/(\pi k_BT)]\}$ occurs~\cite{FVB}.
In what follows we consider the interval of moderately low
temperatures, $E_c/(\pi k_B)<T<\Delta_c/(2k_B)$.

Plugging (\ref{korrfunctlargeT}) into (\ref{current2}), one finally
derives the current-voltage characteristics of the insulating state
as given by Eq.(\ref{IV}). At low biases, $V\ll \Delta_c$, it gives
an Arrhenius activation temperature dependence (\ref{activation})
for the resistance, with the activation energy, $\Delta_c$, defined
by Eq.\,(\ref{Coulombgap}). Carrying out calculations one finds for
a regular array:
\begin{equation}
\Delta_c =
\begin{cases}
(E_c/2)\min{\{\lambda_c,L \}}/d, & \text{for 1D arrays,} \\
(E_c/\pi) \ln (\min{\{\lambda_c,L}\}/d), & \text{for 2D arrays,}
\label{DisordAver2}
\end{cases}
\end{equation}
where $\lambda_c~\simeq d \sqrt{E_{c0}/E_c}$ is the screening length
related to capacitance to ground~\cite{notelambda}. Usually, in
experiments $E_{c0}$ is so high that $\lambda_c$ well exceeds the
sample size $L$.

\begin{figure}[b]
\includegraphics[width=3.3in]{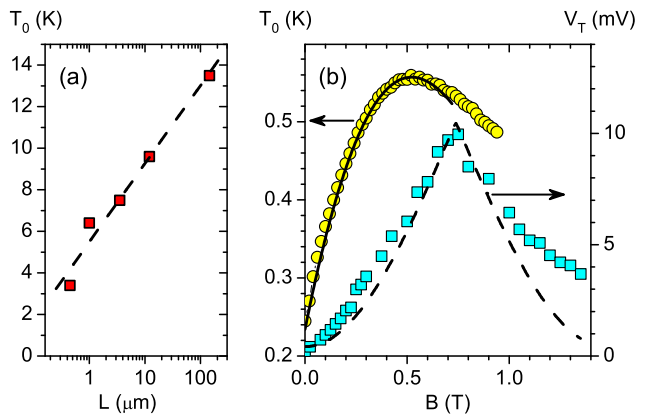}
\caption{(Color online) (a) Activation energy
$T_0\equiv\Delta_c/k_B$ plotted as function of logarithm of the
sample length (squares are the data from~\cite{Ovadyahu}).  (b) The
experimental data from~\cite{Baturina}: activation energy (circles,
left axis) and voltage threshold (squares, right axis) as functions
of the magnetic field $B$. The lines are are according to
Eq.~(\ref{Magnfield}): $E_J^{2D}(B)$, with values $\alpha
E_{J0}/E_c=0.8$ and $A_{loop}=1.4 \cdot 10^{-3}~\mu$m$^2$ fits
$T_0$ (solid line); $V_T(B)$ is the 1D quantity and is fitted by
$E_J^{1D}$ with \textit{the same} $A_{loop}$ and $\tilde\alpha
E_{J0}/E_c=0.96$ (dashed line) reflecting slightly different
geometric factor.} \label{activation-log}
\end{figure}

Dependence of the activation energy on the sample size was recently
observed in InO superconducting films~\cite{Ovadyahu}. Plotted in
Fig.~\ref{activation-log}a are activation energies,
$T_0\equiv\Delta_c/k_B$, extracted from Fig.~4a of~\cite{Ovadyahu},
vs. $\log L$. One clearly sees logarithmic scaling of $T_0$ as Eq.
(\ref{DisordAver2}) predicts.

The nature of the Coulomb barrier $\Delta_c$ and its size scaling
can be understood in terms of the mutual phase locking or phase
synchronization in the Josephson junction array. In the Coulomb
blockade regime the charge at each junction is fixed, and, therefore
conjugated phases fluctuate freely.  Yet, the exponentially small dc
Josephson current couples phases of the adjacent junctions to
provide a minimal power dissipation in the array.  This establishes
a global phase-coherent state, and transport occurs as a
\textit{co-tunneling} of Cooper pairs through the whole array. The
probability of such a process in an 1D array is proportional to
$[\exp(-E_c/k_BT)]^N$, giving the total Coulomb barrier as
$\Delta_c\simeq E_cN$.  Another way of thinking is to say that
synchronization builds on the large screening length $\lambda_c$
which allows for the small charge fluctuations at each junction to
interact over the whole system.  In this sense the linear
(logarithmic) scaling of $\Delta_c$ reflects linear (logarithmic)
growth of the Coulomb energy in 1D (2D, respectively) systems. As a
result, synchronization is rigid with respect to disorder: even
large (of the order of the quantity itself, but Gaussian)
fluctuations in $E_c$, $E_{c0}$, and $E_j$, as well as the effect of
the offset charges is negligible as compared to the huge magnitude
of $\Delta_c$.  That is why this scaling of $\Delta_c$ holds even in
the amorphous superconducting films~\cite{Ovadyahu}, where the
granularity is of a self-induced
nature~\cite{KowalOvadyahu,Baturina} and the variations in Josephson
coupling strength are small.

The current-voltage characteristic of Eq.(\ref{IV}) is valid as long
as $(\Delta_c-eV)^2\gg 2 \Delta_c k_BT $. At temperatures of
interest, $T<\Delta_c/(2k_B)$, this gives an accurate estimate for
the threshold voltage of the \textit{regular} Josephson junction
array as: \vspace{-.1in}
\begin{equation}
eV_T\simeq \Delta_c \label{regthrshld}\, ,
\vspace{-.1in}
\end{equation}
with $\Delta_c$ from~(\ref{DisordAver2}). This result
holds in disordered systems as well.  However, in both 1D and 2D
systems the threshold voltage scales linearly with the size of the
sample, i.e. the 1D scenario works. Consider first an 1D chain. The
current state forms at the threshold voltage as a result of the
dielectric breakdown where the \textit{collective} charge transfer
over the whole array occurs. The associated Coulomb energy
$\Delta_c\sim (L/d)\langle E_c\rangle$ again scales linearly with
the system size as long as the distribution in $E_c$ is not
exponentially broad and the average $\langle E_c\rangle$ is well
defined. The size dependence of $V_T$ on the sample size was
observed in~\cite{Haviland}, where the chain of SQUIDs,
schematically shown in the upper panel of Fig. 1, was studied.  One
can see from the Fig. 2d of~\cite{Haviland} that for two largest
samples indeed $V_T\propto L$. In 2D arrays with disorder the
dielectric breakdown becomes of a percolative nature and occurs
along the first ``lowest resistance" path connecting the leads. This
retains a 1D scaling of $V_T$. Consequently, one expects that in 2D
films the corresponding energy $eV_T$ is much larger than the
activation energy determined from the low bias resistance behavior
(\ref{activation}). Indeed observed in~\cite{Baturina} was
$eV_T/k_BT_0\approx 220$ at the magnetic field 0.7\,T. The above
percolative picture is identical to the threshold charge depinning
in 2D arrays of metallic dots investigated numerically
in~\cite{Middleton}.

Next, we discuss the effect of the magnetic field on the activation
energy and voltage depinning threshold.  The field modulates the
effective Josephson coupling: in 1D SQUID chain one has
$E_J^{1D}=E_{J0}|\cos(\pi f)|$, while in the 2D array
$E_J^{2D}=E_{J0}\{1-4f\sin^2[\pi(1-f)/4]\}$~\cite{Tinkham83}, where
$f = eBA_{loop}/\pi h$, $A_{loop}$ is the area of either the
elemental SQUID or the plaquette in the 2D array. The correction to
Coulomb barrier in the first order perturbation theory with respect
to $E_J/E_c$ follows from~(\ref{korrfunctlargeT},\ref{Coulombgap}):
\begin{equation}
\Delta_c(B)=\Delta_c[1-\alpha E_J(B)/E_c]~,
 \label{Magnfield}
\end{equation}
where the parameter $\alpha$ is of the order of unity, and depends
on the geometry of the lattice. The field modulation of $E_J(B)$
yield non-monotonic field behaviors of $T_0$ and $V_T(B)$. Shown in
Fig.\,2b are fits to activation energy $T_0$ and $V_T$ vs $B$
dependencies to the experimental data from~\cite{Baturina}. The
quantity $\alpha E_{J0}/E_c=0.8$ is chosen to match $T_0(0)$ to
$B=0$ experimental value and reflects that experiments were carried
out in the vicinity of SIT (still allowing ``borderline" estimates
within the perturbation theory). The loop area is defined
unambiguously by the position of the maximum in $T_0$ (only the
branch corresponding $0\leq f\leq 1/2$ should be taken in
$E_J^{2D}$~\cite{Tinkham83}). Using the \textit{same} $A_{loop}$,
the theoretical $V_T(B)$ matches with the data of~\cite{Baturina}
(Fig. 2b); this procedure means the effective absence of fitting
parameters. Fig. 2 confirms the 2D nature of activation energy and
the 1D scenario of depinning threshold.

In conclusion, we have developed a theory of collective Cooper pair
transport in the insulating state of one- and two-dimensional
Josephson junction arrays. We have obtained the Arrhenius low-bias
resistance and derived the corresponding activation energy.  We have
shown that both, the activation energy and the voltage depinning
threshold, represent the dynamic Coulomb barrier $\Delta_c$
controlling collective charge transfer in the insulating state. In
Josephson junction chains the activation energy and voltage
threshold coincide and both scale linearly with the chain length. In
two-dimensional arrays the activation energy scales logarithmically
with the sample length, while threshold voltage, $V_T$, exhibits the
1D linear scaling, since disorder sets the percolative dielectric
breakdown mechanism of charge depinning. We have proposed that the
physical origin of the energy gap and its scaling is the mutual
phase-locking in junction arrays which maintains even in disordered
systems. We expect that at temperatures above the energy gap
$\Delta_c/k_B$ the coherent synchronized state breaks down and the
collective activation transport transforms into a usual local
variable range hopping as observed in \cite{KowalOvadyahu,Baturina}.
We have demonstrated that  modulating Josephson coupling by the
magnetic field leads to a peaked $V_T(B)$-dependence in agreement
with the experimental findings for 1D Josephson
arrays~\cite{Haviland} and for superconducting films near
SIT~\cite{Shahar-Coll,Baturina}.

We like to thank K. B. Efetov and Yu. Galperin for useful
discussions, and Z. Ovadyahu for enlightening conversation and
sharing with us his results prior to publication. V.V. and T.B. are
grateful to V. Kravtsov and M. Kiselev for hospitality at ICTP
Trieste, where the part of this work was completed. This work was
supported by the U.S. Department of Energy Office of Science through
contract No. DE-AC02-06CH11357, SFB 491 and Alexander von Humboldt
Foundation (Germany), RFBR Grant No.\,06-02-16704 and the Program
``Quantum Macrophysics'' of the Russian Academy of Sciences.


\begin{thebibliography}{99}
\bibitem{note1}In the
limit of small tunneling conductance, $g\ll 1$, the junction
charging energy is $E_c=(2e)^2/(2C)$, where $C$ is the junction
capacitance. At $g\gtrsim 1$, $E_c$ is renormalized by the dynamic
screeing effects, $E_c\rightarrow\Delta/g$~\cite{Vinokur}.
\end{thebibliography}
\end{document}